\documentclass[pre,twocolumn,floatfix,bibnotes]{revtex4}

\usepackage{here}
\usepackage[dvips]{graphicx}

\newcommand\pictc[5]{\begin{figure}
                       \centerline{
                       \includegraphics[width=#1\columnwidth]{#3}}
                   \protect\caption{\protect\label{fig:#4} #5}
                    \end{figure}            }
\newcommand\pict[4][1.]{\pictc{#1}{!tb}{#2}{#3}{#4}}
\newcommand\rpict[1]{\ref{fig:#1}}
\newcommand\leqt[1]{\protect\label{eq:#1}}
\newcommand\reqtn[1]{\ref{eq:#1}}
\newcommand\reqt[1]{(\reqtn{#1})}

\newcommand{\be}{\begin{equation}}
\newcommand{\ee}{\end{equation}}
\newcommand{\bea}{\begin{eqnarray}}
\newcommand{\eea}{\end{eqnarray}}

\newcounter{Fig}

\begin{document}

\begin{sloppy}

\title{Second-harmonic generation in nonlinear left-handed metamaterials}

\author{Ilya V. Shadrivov$^1$, Alexander A. Zharov$^{1,2}$, and Yuri S. Kivshar$^1$}

\affiliation{$^1$ Nonlinear Physics Center, Research School of
Physical Sciences and Engineering, Australian National
University, Canberra ACT 0200, Australia \\
$^2$ Institute for Physics of Microstructures, Russian Academy of
Sciences, Nizhny Novgorod 603950, Russia}

\begin{abstract}
We study the second-harmonic generation in left-handed
metamaterials with a quadratic nonlinear response. We demonstrate a novel
type of the exact phase matching between the backward propagating wave of the fundamental frequency and the forward propagating wave of the second harmonics. We show that this novel parametric process can convert a surface of the
left-handed metamaterial into an effective mirror totally
reflecting the second harmonics generated by an incident wave. We
derive and analyze theoretically the coupled-mode equations for a
semi-infinite nonlinear metamaterial. We also study numerically
the second-harmonic generation by a metamaterial slab of a finite
thickness, and reveal the existence of multistable nonlinear effects.
\end{abstract}

\maketitle

\section{Introduction}

Recent years demonstrated many advances in the design and
engineering of artificial composite structures with unique
electromagnetic response, which have broadened significantly the
range of possible wave phenomena not usually observed in Nature
but being accessed in laboratory experiment. In particular, it has
been shown that the composite metallic periodic structures may
allow realizing hypothetical and surprising materials, first
predicted theoretically and termed {\em left-handed
media}~\cite{Veselago:1967-2854:SPSS}, which possess
simultaneously negative dielectric permittivity and magnetic
permeability. Indeed, the composite metallic structures consisting
of arrays of wires and split-ring-resonators have been
demonstrated to possess left-handed properties in the microwave
frequency range~\cite{Smith:2000-4184:PRL}. These composite
materials received a great deal of interest due to their highly
nontrivial and quite often counterintuitive electromagnetic
properties. In particular, such materials support backward waves,
and they exhibit negative refraction at interfaces~\cite{Shelby:2001-77:SCI}.

The study of linear wave propagation and linear properties of
left-handed materials is a major subject of research in this
field. It is indeed the case when both magnetic permeability
and dielectric permittivity of the material do not depend on
the intensity of the electromagnetic field. However, the future
efforts creating {\em tunable structures} where the field
intensity changes the transmission properties of the composite
structure would require the knowledge of {\em nonlinear properties} of
such metamaterials, which may be quite unusual. Indeed, inserting
small conducting elements into a periodic metallic structure of
the metamaterial brings an unique opportunity to enhance nonlinear response in such left-handed media because the microscopic electromagnetic field inside the structure resonant elements can become much stronger than an average
macroscopic field. Indeed, it has already been shown analytically
and numerically that the composite material containing nonlinear
elements in the slits of split-ring resonators possesses a
hysteresis-type nonlinear magnetic
response~\cite{Zharov:2003-37401:PRL,Zharova:2005-1291:OE}. This
is an example of self-action of electromagnetic waves usually
associated with the cubic nonlinearity or its different
generalizations.

Inclusion the elements with non-symmetric current-voltage
characteristics such as diodes into the split-ring resonators will
result in {\em a quadratic nonlinear response} of the
metamaterial~\cite{Lapine:2003-65601:PRE}. This quadratic
nonlinearity is responsible for the recently analyzed parametric
processes such as the second-harmonic generation
(SHG)~\cite{Agranovich:2004-165112:PRB} and three-wave
mixing~\cite{Lapine:2004-66601:PRE}. In particular, the first
analysis of SHG from a semi-infinite left-handed medium has been
briefly presented by Agranovich {\em et
al.}~\cite{Agranovich:2004-165112:PRB}, who employed the nonlinear
optics approach and also made a statement that for
metamaterials {\em ``... SHG in transmission is badly
phase-mismatched''} parametric process, and {\em ``it is then the
SHG in reflection that is more interesting''.} (see
Ref.~\cite{Agranovich:2004-165112:PRB}, p. 165112-5).

In this paper we consider the problem of SHG during the scattering from a semi-infinite left-handed medium (or a slab of the left-handed material of a finite extent) and demonstrate that the original paper by Agranovich {\em et
al.}~\cite{Agranovich:2004-165112:PRB} missed an important
additional phase-matching condition, quite specific for the
harmonic generation by the backward waves. With this condition, we
demonstrate that exact phase matching between a backward
propagating wave of the fundamental frequency (FF) and the forward
propagating wave at the second harmonics (SH) is indeed possible.
This novel phase matched process allows for creating an effective
``quadratic mirror'' that reflects the SH component
generated by an incident FF wave.

The paper is organized as follows. In Sec. 2 we describe 
our model including both the electric and magnetic responses.
Quadratic nonlinearity and the SHG process in metamaterials are
discussed in Sec. 3. In Sec. 4, we develop the corresponding coupled-mode theory for SHG with backward waves and present the analysis of both lossy and lossless
cases of this model. Section 5 is devoted to the case
of SHG process a slab of finite-extension. At last, Sec. 6
concludes the paper.

\section{Model}

We consider a three-dimensional composite structure consisting of
a cubic lattice of conducting wires and
split-ring resonators (SRR), shown schematically in the insert of
Fig.~\rpict{frequency_domains}. We assume that the unit-cell size of the
structure $d$ is much smaller then the wavelength of the propagating
electromagnetic field and, for simplicity, we choose a
single-ring geometry of the lattice of SRRs. The results
obtained for this case are qualitatively similar to those obtained
in more involved cases of double SRRs. This type of
microstructured medium is known to possess the basic properties
of left-handed metamaterials exhibiting negative refraction in the
microwave region.

In the effective-medium approximation, a response of this composite metallic
structure can be described by averaged equations allowing to
introduce the effective dielectric permittivity and effective
magnetic permeability of the form
\begin{equation}\leqt{epsilon}
\epsilon (\omega) = 1 - \frac{\omega_p^2}{\omega^2},
\end{equation}
\begin{equation}\leqt{mu}
\mu (\omega) = 1 + \frac{F\omega}{(\omega_0^2 - \omega^2)},
\end{equation}
where $\omega_p$ is the effective plasma frequency, $\omega_0$ is
a resonant frequency of the array of SRRs, $F$ is the form-factor
of the lattice, and $\omega$ is the angular frequency of the
electromagnetic waves. The product of permittivity $\epsilon$ and
permeability $\mu$ defines the square of the effective refractive index, $n^2 =
\epsilon \mu$, and its sign determines if waves can ($n^2>0$) or
cannot ($n^2<0$) propagate in the medium. Due to the medium
dispersion defined by the dependencies (\ref{eq:epsilon}) and
(\ref{eq:mu}), the wave propagation becomes possible only in
certain frequency domains while the waves decay for other frequencies.
Metamaterial possesses left-handed properties when both $\epsilon$
and $\mu$ become simultaneously negative, and such a frequency domain
exists in the model described by Eqs.~(\ref{eq:epsilon}) and (\ref{eq:mu}) provided $\omega_p
> \omega_0$. In this case, the metamaterial is left-handed within
the frequency range
\begin{equation}
\leqt{range}
 \omega_0 < \omega < {\rm min}
\left\{\omega_p, \omega_M \right\}, \;\; \omega_M =
\frac{\omega_0}{\sqrt{1-F}},
\end{equation}
where $\omega_p$ is the plasma frequency introduced in
Eq.~(\ref{eq:epsilon}).

We assume that $\omega_M < \omega_p$, and in this case we have two
frequency ranges where the material is transparent, the range
where the material is left-handed (LHM), and the right-handed (RHM)
domain for $\omega > \omega_p$, where both permittivity and
permeability are positive (shaded domains in
Fig.~\rpict{frequency_domains}). For the frequencies outside these
two domains, the composite material is opaque.

\pict{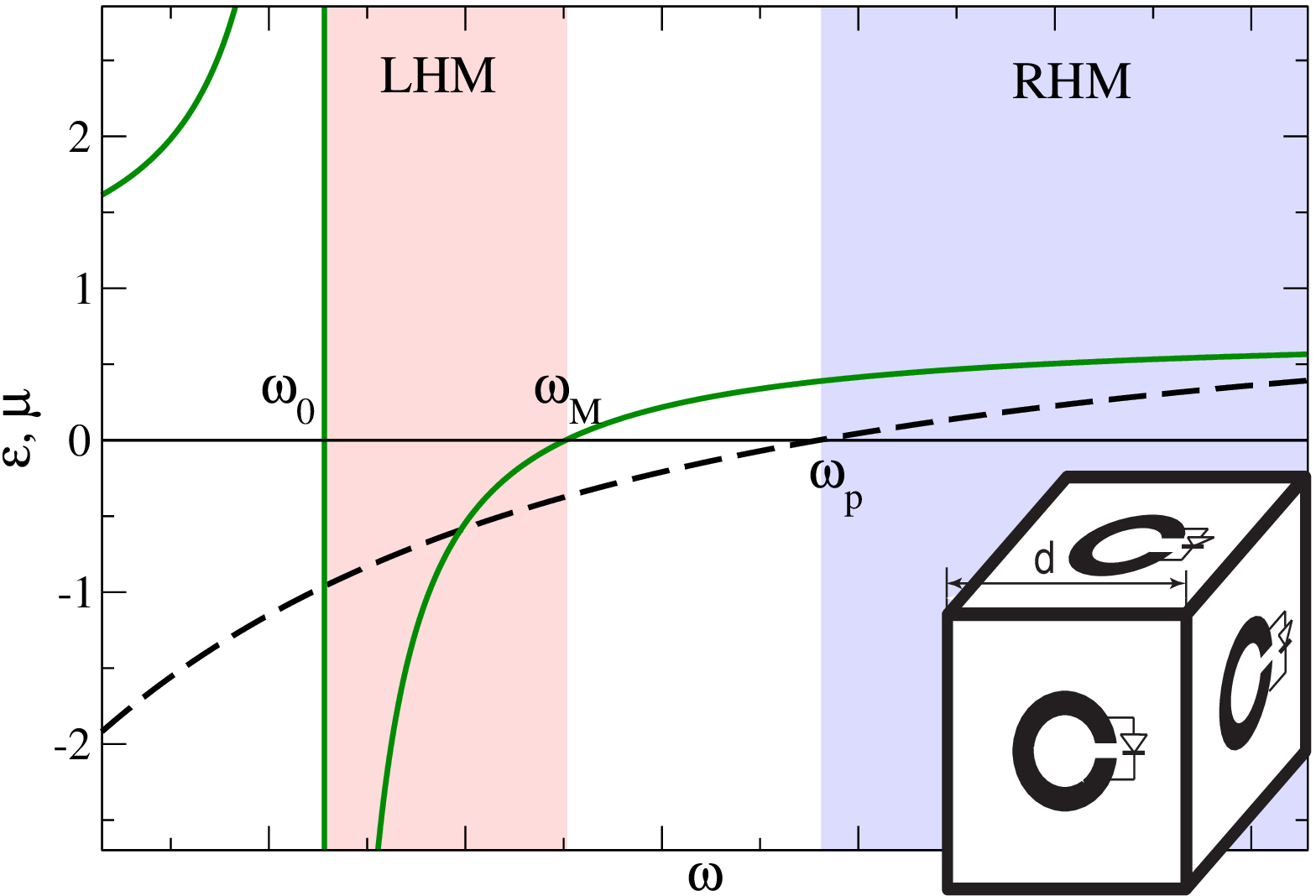}{frequency_domains}{Frequency-dependent
magnetic permeability $\mu$ (solid) and electric permittivity $\epsilon$ (dashed) of the composite. Two types of the regions (LHM or RHM) where the
material is transparent are shaded. For other frequencies it is
opaque. Characteristic frequencies $\omega_0$, $\omega_M$, and
$\omega_p$ are defined in Eqs.~ (\ref{eq:epsilon}) to
(\ref{eq:range}). Inset shows the unit cell of the metamaterial.}

\section{Quadratic nonlinearity and basic equations}

The composite material becomes nonlinear and it possesses a quadratic
nonlinear response when, for example, additional diodes are
inserted into the SRRs of the
structure~\cite{Lapine:2003-65601:PRE}, as shown schematically in
the insert of Fig.~\rpict{frequency_domains}. Quadratic nonlinearity is known
to be responsible for various parametric processes in nonlinear
media, including the frequency doubling and generation of the
second-harmonic field. In dispersive materials, and especially in the
metamaterials with the frequency domains with different
wave properties, the SHG process can be rather nontrivial because
the wave at the fundamental frequency and the second
harmonics can fall into {\em different domains} of the material
properties.

The most unusual harmonic generation and other parametric processes are
expected when one of the waves (either FF or SH wave) has the
frequency for which the metamaterial becomes left-handed. The specific interest to this kind of parametric processes is due to the fact that the waves in the
left-handed media are {\em backward}, i.e. the energy propagates
in the direction opposite to that of the wave vector. Both
phase-matching condition and nonlinear interaction of the forward
and backward waves may become quite nontrivial, as is known from
the physics of surface waves in plasmas~\cite{Zharov:1991-20:FPL}.
In this paper, we are interested in this type of parametric wave
interactions.

In nonlinear quadratic composite metamaterials, interaction of the
forward and backward waves of different harmonics takes place when
the material is left-handed either for the frequency $\omega$ or the double frequency $2\omega$. Under this condition, there exist two types of the most
interesting SHG parametric processes in metamaterials.

{\em Case I.} The frequency of the FF wave is in the range
$\omega_0/2 < \omega < \omega_M/2$ and, therefore, the SH wave is
generated with the double frequency in the LHM domain (see
Fig.~\rpict{frequency_domains}). For such parameters,
the electromagnetic waves at the FF frequency are non-propagating,
since $\epsilon(\omega)\mu(\omega) < 0$. As a result, the field
with the frequency $\omega$ from this range incident on a semi-infinite left-handed medium will decay exponentially from the surface inside the metamaterial. Taking into account Eqs.~(\ref{eq:epsilon}) and (\ref{eq:mu}), the depth $\delta$ of this skin-layer can be found as
\begin{equation}\leqt{p_depth}
\delta = \left(k_{\parallel}^2 - \epsilon\mu\frac{\omega^2}{c^2}
\right)^{-1/2} <
            \frac{\lambda}{17},
\end{equation}
where $k_{\parallel}$ is the tangential component of the
wavevector of the incident wave, and $\lambda$ is a free space wavelength. For the SH wave generated in this
layer, the metamaterial becomes transparent. In this case, a thin
slab of a metamaterial may operate as {\em a nonlinear left-handed lens}
that will provide an image of the source at the second
harmonics, as suggested recently in Ref.~\cite{arXiv-0404074}.
This case has also been mentioned in the earlier paper on the
second-harmonic generation~\cite{Agranovich:2004-165112:PRB}.

{\em Case II.}  The FF wave is left-handed, whereas the SH wave is
right-handed. Such a process is possible when $\omega_p <
2\omega_0$ (see Fig.~\rpict{frequency_domains}). What is truly
remarkable here is the possibility of exact phase-matching of
the SHG parametric process, in addition to the cases discussed
earlier in Ref.~\cite{Agranovich:2004-165112:PRB}. The
phase-matching conditions for this parametric process are depicted
in the dispersion diagram of Fig.~\rpict{dispersion} for the
propagating waves in the metamaterial where the dispersion of the
plane waves is defined by the relation
\begin{equation}\leqt{dispers}
D(\omega,k) = \left[k^2 -
\epsilon(\omega)\mu(\omega)\frac{\omega^2}{c^2}\right] = 0.
\end{equation}
The exact phase matching takes place when $2k(\omega) =
k(2\omega)$. Different signs of the slopes of the curves at the
frequencies $\omega$ and $2\omega$ indicate that one of the waves
is forward, while the other wave is backward.

\pict{fig02}{dispersion}{Dispersion of plane waves
$k(\omega)$ in the metamaterial. Arrows show the parameters of the
FF and SH waves corresponding to the exact spatio-temporal phase
matching.}

To study the SHG process in metamaterials we consider a composite
structure created by arrays of wires and SRRs. To generate a
nonlinear quadratic response of the metamaterial, we assume that
each SRR contains a diode, as depicted schematically in the inset
of Fig.~\rpict{frequency_domains}. The diode is described by the current-voltage dependence,
\begin{equation}\leqt{diode_IU}
I = \frac{U}{R_d} \left( 1 + \frac{U}{U_c} \right),
\end{equation}
where $U_c$ and $R_d$ are the parameters of the diode, and $U$ is
the voltage on the diode. Equation~\reqt{diode_IU} is valid
provided $U \ll U_c$, and it represents two terms of the Taylor expansion
series of the realistic (and more complex)
current-voltage characteristics of the diode.

Following the standard procedure, we consider two components of
the electromagnetic field at the fundamental frequency $\omega$
and its second harmonics $2\omega$, assuming that all other
components are not phase matched and therefore they give no substantial
contribution into the nonlinear parametric interaction.
Subsequently, we write the general coupled-mode equations
describing the simultaneous propagation of two harmonics in the
dispersive metamaterial as follows,
\begin{eqnarray}\leqt{Helmholtz}
\Delta {\bf H}_1 + \epsilon(\omega)\mu(\omega)\frac{\omega^2}{c^2}
{\bf H}_1 =
    - \sigma_1 {\bf H}_1^* {\bf H}_2, \nonumber \\
\Delta {\bf H}_2 + 4 \epsilon(2\omega)\mu(2\omega)
   \frac{\omega^2}{c^2} {\bf H}_2 = -\sigma_2 {\bf H}_2,
\end{eqnarray}
where the indices ``1'',``2'' denote the FF and SH fields, respectively;
$\Delta$ is a Laplacian, and other parameters are defined as
follows
\bea\leqt{sigm_new} \sigma_1 = \kappa/2R(\omega), \;\;
\sigma_2 = \kappa/R^*(\omega), \nonumber\\
\kappa = \frac{6 \pi \left( \pi a^2 \right)^3 }{d^3 c^5}
    \left[\frac{\omega_0^4\omega^2}{U_c R_d R(\omega) R(2\omega)}\right],
\eea
where $R(\omega) = \omega_0^2 \omega^2 +i\gamma \omega$, the
asterisk stands for the complex conjugation, $a$, $d$ are the
radius of the SRRs and the period of the metamaterial,
respectively, and $\gamma$ is the damping coefficient of the SRR.
For simplicity, we assume that both FF and SH waves are of the
same polarization, and therefore they can be described by only one
component of the magnetic field. In this case,
Eqs.~\reqt{Helmholtz} become scalar. In the derivation of
Eqs.~\reqt{Helmholtz} we take into account the Lorentz-Lorenz
relation between the microscopic and macroscopic magnetic
fields~\cite{Landau:1963:Electrodynamics}. Also, it is assumed
that the diode resistance $R_d$ is much larger than the impedance
of the SRR slit, i.e. $R_d \gg 1/\omega C$, so that the resonant
properties of the composite are preserved.

\section{Semi-infinite metamaterial}

First, we consider a semi-infinite left-handed medium and the SHG
process for the wave scattering at the surface. We assume that a TM-polarized FF wave is incident on a LH material from the vacuum, as shown schematically in
Fig.~\rpict{geom}. Inside the metamaterial, the wave at the
fundamental frequency satisfies the dispersion relation
\reqt{dispers} which defines the wavenumber $k$. As is
discussed above, the FF wave in the left-handed medium is
backward, meaning that the normal component of the wave vector is directed
towards the surface, i.e. in the direction  opposite to the
Poynting vector.

\pict{fig03}{geom}{Geometry of the SHG problem. Thick arrows show the
direction of the energy flow, thin arrows -- direction of wavevectors. Indices ``i'', ``r'', and ``tr'' stand for incident, reflected and transmitted waves, respectively.}

When the phase-matching conditions are satisfied, the generated SH
wave has the wavevector parallel to that of the FF wave (see
Fig.~\rpict{geom}). However, the SH wave is forward propagating,
so that the energy at this frequency should propagate towards the
interface. When losses are negligible, the FF wave will be {\em
transformed completely} into the SH wave with the energy flows in
the direction opposite to that of the FF wave. This kind of the
SHG process in a semi-infinite left-handed medium is characterized
by two major features: (i) the efficiency of the SHG process may
become very high, and (ii) the SH wave propagates in the direction
 opposite to that of the incoming FF wave.

\subsection{Coupled-mode equations}

To describe the SHG process analytically, we employ the
coupled-mode theory and the slowly-varying envelope approximation
for the FF and SH fields, and present the magnetic fields in the
material in the form:
\begin{equation}\leqt{slow_fields}
H_{1,2}(t,z) = a_{1,2}(t,z) e^{ik_{1,2} z} + c.c.,
\end{equation}
where the amplitudes of the FF and SH fields $a_{1,2}(t,z)$ are
assumed to vary slow in both space and time, i.e.  $\partial
a_{1,2}/\partial t \ll \omega a_{1,2}$, and $\partial
a_{1,2}/\partial z  \ll k a_{1,2}$. Substituting
Eqs.~\reqt{slow_fields} into Eqs.~\reqt{Helmholtz} and neglecting
the second-order derivatives, we obtain the coupled equations,
\begin{eqnarray}\leqt{slow_eqn}
\frac{\partial a_1}{\partial t} + v_{g1}\frac{\partial a_1}{\partial z} =
    i\sigma_1 a_1^* a_2 - \nu_1 a_1, \nonumber\\
\frac{\partial a_2}{\partial t} + v_{g2}\frac{\partial a_2}{\partial z} =
    i\sigma_2 a_1^2 + \nu_2 a_2 - i \Omega a_2,
\end{eqnarray}
where $v_{g1,2}$ are the group velocities and $\nu_{1,2} =
v_{g1,2}Im(k)$ are linear damping coefficients of the FF and SH
fields, respectively,
\begin{equation}
\Omega = q_2 D(2\omega, 2k)/2\sigma_2,
\end{equation}
is the phase mismatch, and
\begin{eqnarray}\leqt{sigma}
q_1 = \sigma_1 \left[\frac{\partial
D(\omega,k)}{\partial\omega}\right]^{-1}, \;\;\; q_2 = \sigma_2
\left[\frac{\partial D(2\omega,2k)}{\partial\omega}\right]^{-1}.
\end{eqnarray}
The coupled-mode equations~\reqt{slow_eqn} can be presented in the
equivalent rescaled form,
\begin{eqnarray}\leqt{slow_eqn1}
\frac{\partial b_1}{\partial t} + v_{g1}\frac{\partial b_1}{\partial z} =
    -q_1 b_1^* b_2 - \nu_1 b_1, \nonumber\\
\frac{\partial b_2}{\partial t} + v_{g2}\frac{\partial b_2}{\partial z} =
    q_2 b_1^2 + \nu_2 b_2 - i \Omega b_2,
\end{eqnarray}
where $a_1 = \alpha b_1$, $a_2 = \beta b_2$, $\alpha =
\exp(i\phi)$, $\beta = \exp[-i(\pi/2 - 2\phi)]$, and $\phi$ is
an arbitrary phase.

The incoming FF backward travelling wave has the group velocity in
the $z$-direction, and the phase velocity -- in the opposite direction. The generated SH forward wave has both the phase and group velocities in the $-z$-direction. The FF wave propagates inside the material, and it loses the energy due to SHG and also due to losses in the medium. As a result, the SH amplitude decreases in the $z$-direction, and the boundary conditions should be taken in the
form $b_{1,2}(\infty) = 0$.

\subsection{Lossless process}

First, we neglect the metamaterial losses assuming, for
simplicity, that $\nu_{1,2} = 0$ and $Im(q_{1,2}) = 0$. At the
large enough value of the phase mismatch we can neglect the derivatives in the second equation of the system~\reqt{slow_eqn1}, and obtain a local coupling between the FF and SH amplitudes,
\begin{equation}\leqt{local}
b_2 = -i \frac{q_2}{\Omega}b_1^2.
\end{equation}
Substituting Eq.~\reqt{local} into the equation for the FF field, we obtain a single nonlinear equation for the amplitude $b_1$ with
an effective cubic nonlinearity,
\begin{equation}
\leqt{cubic}
\frac{\partial b_1}{\partial t} + v_{g1}\frac{\partial b_1}{\partial z} =
    i \frac{q_1q_2}{\Omega} |b_1|^2 b_1.
\end{equation}
The general solution of Eq.~\reqt{cubic} can be found in the form
\begin{equation}
\leqt{cubic_solution}
b_1 = A(t, z) e^{i\Phi(t,z)},
\end{equation}
where $A(t,z) = f(t-z/v_{g1})$, $\Phi(t,z) = (t-z/a)q_1 q_2
f^3(t-z/v_{g1})/\Omega(1-v_{g1}/a)$, where $f(t,z)$ is an
arbitrary function, $a$ is constant, and $a\ne v_{g1}$. The
obtained solution describes {\em a stationary nonlinear wave}. 

In the case of the exact phase matching, i.e. when $\Omega = 0$,
Eqs.~\reqt{slow_eqn1} are simplified,
\begin{eqnarray}\leqt{slow_eqn_matching}
\frac{\partial b_1}{\partial t} + v_{g1}\frac{\partial b_1}{\partial z} =
    -q_1 b_1^* b_2, \nonumber\\
\frac{\partial b_2}{\partial t} - v_{g2}\frac{\partial b_2}{\partial z} =
    q_2 b_1^2.
\end{eqnarray}
Looking for stationary solutions of Eqs.~\reqt{slow_eqn_matching},
we take $\partial /\partial t = 0$ and find the integral of motion of
this system,
\begin{equation}\leqt{integral}
\frac{v_{g1}}{q_1}|b_1|^2 - \frac{v_{g2}}{q_2}|b_2|^2 = C.
\end{equation}
Since both the waves should vanish inside the metamaterial, we take $C
= 0$. Then, the solution of Eqs.~\reqt{slow_eqn_matching} is found as
\begin{eqnarray}\leqt{solution}
b_1 = \left(\frac{v_{g1}v_{g2}}{q_1 q_2}\right)^{1/2}
\frac{1}{(z+z_0)}, \;\;\;  b_2 = \frac{v_{g1}}{q_1}
\frac{1}{(z+z_0)},
\end{eqnarray}
where $z_0$ is constant. Solution ~\reqt{solution} has a
singularity, and it corresponds to an explosive instability of the
parametrically interacting waves. However, this singularity does
not appear in the corresponding boundary problem.

\pict{fig04}{slab_scattering}{Geometry of the SHG
process for a finite-width slab of a nonlinear quadratic
metamaterial (shaded). }

Equations~\reqt{slow_eqn_matching} describe also the solution for the
coupled FF and SH fields, where the FF wave has a localized
profile, while the SH wave has a $\rm tanh$ profile,
\begin{eqnarray}
\leqt{kink-soliton}
b_1(t,z) = A_1 {\rm sech}\left[\left( t-z/\beta \right) /T \right],\nonumber\\
b_2(t,z) = A_2 {\rm tanh}\left[\left( t-z/\beta \right) /T \right],
\end{eqnarray}
where %
\begin{equation}
\leqt{solu2}
 A_1 = \left[\frac{(v_{g2}+\beta)(v_{g1}-\beta)}{\beta^2 T^2
q_1 q_2}\right]^{1/2}, \;\; A_2 = \frac{(v_{g1}-\beta)}{\beta T q_1},
\end{equation}
$\beta$ is the pulse velocity, and $T$ is a characteristic time
scale. Solutions \reqt{kink-soliton} are valid
provided $\beta < -v_{g2}$, which means that in the moving reference
frame both waves should be of the same type: either forward
or backward. The analytical solution
~\reqt{kink-soliton} and ~\reqt{solu2} is characterized by two
arbitrary constants, $\beta$ and $T$, so that the corresponding family
of solutions is two-parametric.

\subsection{Dissipative process}

Next, we consider more realistic case with a linear damping of the SH wave smaller than that of the FF wave, since the FF wave is much closer to the resonance where such losses may become quite essential. We assume again $Im(q_{1,2}) = 0$, so that
the stationary solution of Eqs.~\reqt{slow_eqn1} can
be found in this case analytically under the conditions of the exact phase
matching. First, we find the integral of motion of the coupled system
in the form,
\begin{equation}\leqt{integral2}
\frac{v_{g2}}{v_{g1}}\left[ q_1 b_2^2 + 2\nu_1 b_2 \right] = q_2 b_1^2.
\end{equation}
Using this invariant, we find the solutions for the amplitudes of
FF and SH waves as follows
\begin{eqnarray}\leqt{solution_dissipative}
b_1(z) = \nu_1\left(\frac{v_{g2}}{q_1 q_2 v_{g1}}\right)^{1/2}{\rm
csch}
    {\left[ \frac{\nu_1}{v_{g1}}(z+z_0) \right]}, \nonumber\\
b_2(z) = \frac{2\nu_1}{q_1} \left[ e^{2\nu_1(z+z_0)/v_{g1}} -1
\right]^{-1}.
\end{eqnarray}
In the lossless case, Eqs.~\reqt{solution_dissipative} coincide
with Eqs.~\reqt{solution}, however, the presence of finite losses
changes the type of solutions from power to exponential ones.

\pict{fig05}{coeff_vs_input2}{Reflection coefficients of
the generated SH wave (solid) and transmitted FF (dotted), as well as
the transmission coefficient of the FF wave (dashed) vs.
the normalized field intensity. Normalized slab thickness $L=10$.}

\section{A slab of metamaterial}

Next, we study the SHG process for a layer of the thickness $L$
(see Fig.~\rpict{slab_scattering}) and employ a direct numerical
approach to solve Eqs.~\reqt{Helmholtz}. First, we rewrite
Eqs.~\reqt{Helmholtz} in the dimensionless form
\bea\leqt{norm_eqs}
\frac{d^2 H_1}{dz^2} +
    \left[ \epsilon(\omega)\mu(\omega) - k_x^2  \right] H_1 =
        -H_{2} H_1^*, \nonumber\\
\frac{d^2 H_{2}}{dz^2} +
    4\left[ \epsilon(2\omega)\mu(2\omega) - k_x^2  \right] H_{2} =
        - Q H_1^2,
\eea
where the magnetic field is normalized by the value
$\omega^2\sigma_1/c^2 |\sigma_1|^2$, $z$ is normalized by the
value $c/\omega$, and $Q = \sigma_1\sigma_2/|\sigma_1|^2$. We
assume that a slab of the left-handed material is illuminated by
the FF wave with the amplitude $H_\omega^{(i)}$, and the SH wave
is generated inside the slab, so that the reflected and
transmitted waves of both the frequencies $\omega$ and $2\omega$
appear (see Fig.~\rpict{slab_scattering}), with the amplitudes
$H_1^{(t)}$, $H_{2}^{(t)}$, $H_1^{(r)}$, $H_{2}^{(r)}$, respectively. Solving the coupled-mode equations
numerically, we present our results for the reflection and
transmission coefficients defined as $R_{\omega, 2\omega} =
H_{1, 2}^{(r)}/H_1^{(i)}$ and $T_{\omega, 2\omega}
= H_{1, 2}^{(t)}/H_1^{(i)}$.

In the calculations presented here we assume that the left-handed
material is lossless, and we consider the normal incidence. We
take the following parameters of the composite: $\omega_0 = 2 \pi
\times 5$ GHz, $\omega_p = 2 \pi \times 7$ s$^{-1}$, $F = 0.3$,
$a=3$mm, $d=6$mm, $U_c R_d = 10^5$ CGS. For such parameters the
exact phase matching takes place at $f_{pm} \approx 5.37$ GHz.
Dependences of the coefficients $R_{2\omega}$, $T_{\omega}$ and
$R_{\omega}$ vs. the amplitude of the incident FF wave are shown
in Fig.~\rpict{coeff_vs_input2}. One can see that the efficiency
of the transformation of the incident FF wave into the reflected
SH wave can be rather high. Larger intensities of the incident
field result in the multistable behavior of the reflection and
transmission coefficients.

Figure~\rpict{coeff_vs_th_2} shows the dependence of the
transmission coefficient of the FF wave and the reflection
coefficient of the SH wave vs. the slab thickness, for a fixed
amplitude of the transmitted wave. We observe multistable behavior of the coefficients for thicker slabs.
\pict{fig06}{coeff_vs_th_2}{Transmission coefficient of
the FF wave (dashed) and reflection coefficient of the SH wave (solid) vs. the slab thickness $L$, for a fixed amplitude of the transmitted wave,
$H_1(L)=10^{-2}$. Right plot shows a blow up of the region depicted by a dashed box.}

\section{Conclusions}

We have presented a comprehensive study of the specific features
of the second-harmonic generation in left-handed metamaterials
with a quadratic nonlinear response. First, we have demonstrated
a possibility of the exact phase matching between backward propagating FF and
forward propagating SH waves. Then, we have developed
an analytical approach based on a novel type of the coupled-mode
equations in order to characterize the process of the harmonic
generation from a semi-infinite left-handed metamaterial. In particular, we have
demonstrated that a surface of the left-handed metamaterial can
operate as an effective mirror that reflects all generated SH waves.
Finally, we have performed numerical simulations of the second-harmonic
generation process for a finite slab of the metamaterial, and revealed quite unusual multi-valued multistable transmissions and reflections.

\section{Acknowledgments}

The work was supported by the Australian Research Council.
Alexander Zharov acknowledges a warm hospitality of the Nonlinear
Physics Centre during his stay in Canberra, as well as a support
from the Russian Fund for Basic Research (grant N05-02-16357).

\end{sloppy}
\end{document}